\documentclass[superscriptaddress,
amsmath,amssymb
pra,twocolumn,
floatfix
]{revtex4-2}

\usepackage{graphicx}
\usepackage{xcolor}
\usepackage{threeparttable}
\usepackage[normalem]{ulem}
\usepackage{tabularx}

\begin{document}

\title{Optical excitation of ultra-relativistic partially stripped ions}

\author{Jacek Biero\'n}
\affiliation{Institute of Theoretical Physics, Jagiellonian University, ul. {\L}ojasiewicza 11, 30-348 Krak\'ow, Poland}
\author{Mieczyslaw Witold Krasny}
\affiliation{LPNHE, Sorbonne University, CNRS/IN2P3, Tour 33, RdC, 4, pl.\ Jussieu, 75005 Paris, France}
\affiliation{CERN, Esplanade des Particules 1, 1211 Geneva 23, Switzerland}
\author{Wies{\l}aw P{\l}aczek}
\affiliation{Institute of Applied Computer Science, Jagiellonian University, ul.\ {\L}ojasiewicza 11, \\ ~~30-348 Krak\'ow, Poland}
\author{Szymon Pustelny}
\email{Email: pustelny@uj.edu.pl}
\affiliation{M. Smoluchowski Institute of Physics, Jagiellonian University, ul. {\L}ojasiewicza 11, 30-348 Krak\'ow, Poland}

\begin{abstract}

The Gamma Factory (GF) initiative aims at the construction of a unique experimental tool exploiting resonant interaction of light with ultra-relativistic partially stripped ions (PSI) stored in circular accelerators at CERN. Resonant excitation of high-energy electronic transitions in the ions is achieved through Doppler-boosting (by twice the Lorentz factor; from hundred to several thousand times) of light energy. In order to efficiently excite the ions, and hence generate intense beams of scattered/fluorescent photons, a detailed knowledge of the ions' electronic energy structure and the dynamics of optical excitation is required. Spectroscopic properties of PSI selected for the GF operation, as well as their optical excitation schemes, are investigated.  Two regimes of the ion--light interaction are identified, leading to different dynamics of the excitation.  The efficiency of the ion--light interaction, as well as the number of photons emitted from a single ion bunch, are estimated, both analytically and numerically, for three ions considered for the GF, i.e.~Li-like ${}^{208}_{\phantom{0}82}$Pb$^{79+}$, Li-like ${}^{40}_{20}$Ca$^{17+}$, and H-like ${}^{208}_{\phantom{0}82}$Pb$^{81+}$.

\end{abstract}

\keywords{Gamma Factory, partially stripped ions, ultra-relativistic ions, optical excitation, resonant absorption, Rabi oscillations, electronic transitions in partially stripped ions}

\maketitle

\section{Introduction}
\label{sec:Intro}

The goal of the Gamma Factory (GF) project \cite{Krasny:2015ffb} is to develop new tools for the CERN-based research programme.  They include  \cite{GF-PoP-LoI:2019,Placzek:2019xpw,GammaFactoryWorkingGroup:2020ely,Budker:2020zer,Krasny:2021llv,Budker:2021fts,Nichita:2021iwa}: (1) atomic traps of highly charged atoms; (2) an electron beam for electron--proton collisions in the LHC interaction points; (3) high-intensity photon beams; (4) laser-light-based cooling methods of high-energy hadronic beams; and (5) high-intensity beams of polarized electrons, polarized positrons, polarized muons,  neutrinos, neutrons, and radioactive ions.  It is the first accelerator-technology-based project for which precise and accurate atomic-physics input is indispensible.  Firstly, to maximize the intensity of the GF beams, a very efficient optical excitation of the highly-ionised atoms, stored in the CERN accelerator complex, is of pivotal importance.  Secondly, this project requires high-precision calculations of the atomic energy levels of hydrogen, helium and lithium-like partially stripped ions (PSI) to precisely tune the ion beam energy to resonantly excite specific electronic transitions with Doppler-boosted laser light.  Since linewidths of the atomic resonant excitations are much narrower than frequencies of the transitions, the ``resonance-finding'' procedure, involving tuning of the ion relativistic Lorentz factor $\gamma_{L}$,  will certainly be one of the most difficult operation aspects of the project.  Thirdly, the lifetimes of the excited states must be accurately calculated in order to optimize and control the direction, polarization and energy of fluorescence photons. 
On top of precise knowledge of the static parameters of the PSI, which will be used in the GF research programme, good understanding of the ion--light interaction dynamics is necessary to optimize the parameters of laser pulses as well as to determine spatiotemporal characteristics of ion bunches.

One of the pivotal milestones of the GF project is the development of a theoretical framework for the interaction of light pulses with ion bunches.  In general, the interaction has to be described as a quantum-mechanical process which includes interference of probability amplitudes.  In this scope, its probabilistic description, which does not take into account the quantum superposition of atomic states, is approximate and can only be used in limited context.  At the same time, implementation of the process in terms of probabilistic (classical) observables is tempting as it enables application of Monte-Carlo generators. The Monte-Carlo framework provides an efficient interface between the theoretical framework and existing software tools, used to describe both the individual particles and collective beam dynamics in high-energy storage rings. 

The basic goal of the present paper, being a crucial step for the development of the GF, is to provide a clear assessment under which conditions the probabilistic framework is sufficient and under which conditions it fails.  These initial conditions will be specified by the following parameters: (1) the laser-pulse temporal and spatial shape, e.g.\ its time-dependent energy spectrum and total energy, and (2) the ion-bunch shape and the geometry of interactions. 

The paper is organized as follows. In Section~\ref{section.IonChoice} we investigate: the properties of the Li-like Pb ion which may be suitable for the proof-of-principle experiment (PoP) in the Super Proton Synchrotron (SPS) at CERN \cite{GF-PoP-LoI:2019}; the H-like Pb ion which is considered as a candidate for the GF experiment at the LHC \cite{Bessonov:1995eq}; and Li-like Ca ions which can be used for a high-luminosity option of the LHC with laser-cooled isoscalar ion beams \cite{Krasny:2020wgx,Krasny:2021llv}. We have also included in this section the He-like Ca ion case because of its potential future use for the production of radioactive ion beams \cite{Nichita:2021iwa}.  For the PoP experiment at the SPS, we need an ion which satisfies the following criteria: (1) transition energy of $10\,$--$\,100\,$eV and (2) lifetime of the excited state of $10\,$--$\,100\,$ps.
We have verified that these conditions are fulfilled, both with respect to the energy and the lifetime, for H-like, He-like, Li-like, Ne-like, Mg-like, Ar-like and Kr-like isoelectronic sequences.  We have not investigated Xe-like or Rn-like, nor any other sequence, but it can be done if necessary.  In Section~\ref{sec:OptPump}, we discuss laser-light excitation of PSI in the GF experiments. The calculations are performed for two distinct scenarios: (1) where the Doppler broadening dominates over the natural linewidth of the transition and (2) where these two widths are comparable. These lead to two different dynamics of light absorption and reemission.
Finally, Monte-Carlo simulation of interactions of a PSI bunch with a laser-light pulse is described in Section~\ref{sec:MCsim}.  The conclusions are presented in Section~\ref{sec:Concl}.

\newcommand{\thicksim}{\,\mathbf{\sim}\,}
\section{Energies and lifetimes of atomic excited levels}
\label{section.IonChoice}

Partially stripped ions which are presently considered as the candidates for the GF are:  Li-like Pb, H-like Pb, He-like Ca, and Li-like Ca.  The energies $ E_{ik} $ and lifetimes $ \tau_{ik} $ for selected transitions in these ions (as well as other candidates mentioned above) were computed with the Dirac--Hartree--Fock package {\sf GRASP}~\cite{compas.github,grasp2K:2013,GRASP2018,GrantBook2007,Grant1994}.  These calculations were made in the Dirac--Hartree--Fock model, i.e.~with the Dirac--Coulomb Hamiltonian, with a finite nuclear size modeled as the two-parameter Fermi distribution~\cite{grasp89} and with the leading QED corrections (SE and VP) evaluated perturbatively~\cite{graspMcKenzie1980}.  The electron correlation effects were accounted for through the multiconfiguration variational Complete Active Space approach~\cite{BieronAu2009,Bieron:e-N:2015}.  The calculated values were compared with other data available in the literature, and in each case the most accurate results were selected in Tables~\ref{table.Li-like-Pb}, \ref{table.H-like-Pb}, \ref{table.He-like-Ca}, \ref{table.Li-like-Ca}, and~\ref{table:OpticalPumpingParameters}.

\subsection{Li-like Pb}
\label{section.Li-like-Pb}
The NIST-ASD database~\cite{NIST-ASD} lists 46 references on the subject of Pb$^{79+}$.  Transition energies and lifetimes for the Li-like~Pb ion are collected in Table~\ref{table.Li-like-Pb}.  For each calculation we quote as many digits as provided by the authors.  Only the three latest calculations include the estimates of the uncertainty.  Finally, in Table~\ref{table:OpticalPumpingParameters} we adopted the 2s--2p$_{1/2}$ transition energy
$ 230.823(47)(4) $~eV in Li-like~Pb from Yerokhin and Surzhykov~\cite{YerokhinSurzhykov:2018a,YerokhinSurzhykov:2018b}, and the lifetime for $ 2p_{1/2} $ state of Li-like Pb from Johnson, Liu, and Sapirstein~\cite{JohnsonLiuSapirstein:1996}.  They estimated that their calculated lifetimes are accurate to a fraction of a percent at the neutral end of the isoelectronic sequence, and the accuracy increases at higher $Z$.

\begin{table*}[htbp!]\centering
    \caption{\normalsize{Transition energies $E$ and lifetimes $\tau$ of the 2s--2p$_{1/2}$ and 2s--2p$_{3/2}$ lines in the Li-like Pb ion.}}
    \vspace{0.1cm}
    \begin{tabular}{llllclc}
        \hline\hline
        \multicolumn{2}{c}{2s--2p$_{1/2}$} & 
        \multicolumn{2}{c}{2s--2p$_{3/2}$} \\
        \hline
        $E$ [eV]       &$\tau$ [ps] &  $E$ [eV]       & $\tau$ [fs] & year & method & \multicolumn{1}{c}{reference} \\
        \hline 
        231.374        && 2642.297        && 1990 & MCDF VP SE & \cite{IndelicatoDesclaux} \\
        230.817        && 2641.980        && 1991 & MCDF VP SE & \cite{Kim} \\
        230.16 & 73.96  & 2649.23 & 41     & 1991 & Coul-App HS-core & \cite{TheodosiouCurtisEl-Mekki:1991} \\ 
        230.698        && 2641.989        && 1995 & RCI QED NucPol & \cite{ChenChengJohnsonSapirstein} \\
        231.16 & 76.6   & 2642.39 & 42.22  & 1996 & 3-rd order MBPT & \cite{JohnsonLiuSapirstein:1996} \\
        230.650(30)(22)(29) &&  ---    && 2003 & expt(DR) & \cite{Brandau:2003} \\
        ---           && 2642.26(10)     && 2008 & expt(EBIT) &  \cite{Zhang} \\
        230.68         &&  ---            && 2010 & RCI QED NucPol & \cite{Kozhedub} \\ 
        230.76(4)      && 2642.17(4)      && 2011 & S-m. 2-l. NucPol & \cite{SapirsteinCheng} \\
        230.823(47)(4) && 2642.220(46)(4) && 2018 & RCI QED NucPol & \cite{YerokhinSurzhykov:2018a,YerokhinSurzhykov:2018b} \\
        230.80(5)      && 2642.20(5)      && 2019 & S-m. 2-l. NucPol & \cite{pcSapirsteinCheng} \\
        232 & 76   & 2643 & 42 & 2021 & RCI VP SE & this work \\       
        \hline \hline
        \vspace{0.1cm}
        \end{tabular}
    \label{table.Li-like-Pb}
\end{table*}

\subsection{H-like Pb}
\label{section.H-like-Pb}

\begin{table}[htbp!]\centering
    \caption{\normalsize{Transition energies $E$ and lifetimes $\tau$ of the 1s--2p$_{1/2}$ and 1s--2p$_{3/2}$ transitions in the H-like Pb ion.}}
    \begin{tabular}{lc|lc|ccc}
        \hline\hline
        \multicolumn{2}{c}{1s--2p$_{1/2}$} & 
        \multicolumn{2}{c}{1s--2p$_{3/2}$}  \\
        \hline
        $E$ [eV]       &$\tau$ [as] & $E$ [eV] & $\tau$ [as] & year & reference \\
        \hline 
        75280.47     & ---  &  77934.25     & ---  & 1985 &  \cite{JohnsonSoff:1985} \\
        75279        & ---  &  ---          & ---  & 1997 &  \cite{Beier:1997} \\
        75521        & 33.8 &  78174        & 38.8 & 2003 &  \cite{JitrikBunge:2003} \\
        75280.83(26) & ---  &  77934.59(27) & ---  & 2015 &  \cite{YerokhinShabaev:2015} \\
        75278        & 34.1 &  77935        & 39.2 & 2021 & this work \\
        \hline\hline
    \end{tabular}
    \label{table.H-like-Pb}
\end{table}

The bibliography of spectroscopic properties of hydrogen-like ions lists more then one hundred papers~\cite{NIST-ASD,YerokhinShabaev:2015}.  In the present work (see Table~\ref{table.H-like-Pb}) we have taken into consideration the papers of Johnson and Soff~\cite{JohnsonSoff:1985}, Beier~\textit{et~al.}~\cite{Beier:1997}, Jitrik and Bunge~\cite{JitrikBunge:2003}, and Yerokhin and Shabaev~\cite{YerokhinShabaev:2015}.  Johnson and Soff~\cite{JohnsonSoff:1985} took into account the QED effects (Lamb shift), the effects of the finite nuclear size, reduced mass, nuclear recoil effects, as well as their respective cross-terms.  Beier~\textit{et~al.}~\cite{Beier:1997} took into account the QED effects, the effect of the finite nuclear size, as well as the nuclear recoil effect.  Jitrik and Bunge~\cite{JitrikBunge:2003} evaluated the transition energies and rates for hydrogen-like ions using eigenfunctions of the Dirac Hamiltonian with a point nucleus.  The discrepancy between the values of Jitrik and Bunge~\cite{JitrikBunge:2003} and the values obtained with more elaborate approaches of Johnson and Soff~\cite{JohnsonSoff:1985}, Beier~\textit{et~al.}~\cite{Beier:1997}, and Yerokhin and Shabaev~\cite{YerokhinShabaev:2015} illustrates the contributions of the effects beyond the
the Dirac Hamiltonian with a point nucleus.  Yerokhin and Shabaev~\cite{YerokhinShabaev:2015} took into account the QED effects, the finite nuclear size, the nuclear recoil as well as their respective cross-terms.  In particular, they thoroughly evaluated the two-loop QED correction and the finite nuclear size correction, which constitute the dominant sources of uncertainty for the H-like Pb ion.  All these authors adopted different values of the fine structure constant $\alpha$ which were considered standard at the respective publication dates.  These differences, ranging between seventh up to tenth figure (the current value of \mbox{$\alpha^{-1}$~=~137.035~999~084(21)}~\cite{NIST-Constants}) contributed to several factors involved in the summation of the transition energy and rate.  In Table~\ref{table:OpticalPumpingParameters} we adopted the energy for the transition \mbox{$E(1\mathrm{s}$--$2\mathrm{p}_{1/2})$~=~$75280.83(26)$~eV} in the H-like~Pb ion from Yerokhin and Shabaev~\cite{YerokhinShabaev:2015} and the lifetime (34 attoseconds) calculated in the present work with the Dirac--Coulomb Hamiltonian, with the finite nuclear size and with the leading QED corrections evaluated perturbatively~\cite{graspMcKenzie1980}.
 
\subsection{He-like Ca}
\label{section.He-like-Ca}
The calculations of the transition energies and rates for the He-like Ca ion were performed with the {\sf GRASP} package described at the beginning of Section~\ref{section.IonChoice}.  The results are presented in Table~\ref{table.He-like-Ca} and compared with data available in the literature.  The transition energy calculated by Artemyev~\textit{et~al.}~\cite{Artemyev:2005} is the most reliable among those presented in 
Table~\ref{table.He-like-Ca}.  For the lifetime, one might cautiously adopt $ 6.0(1)$\,ps, i.e.\ an average of the three values calculated by Lin~\textit{et~al.}~\cite{LinJohnsonDalgarno:1977}, Aggarwal~\textit{et~al.}~\cite{Aggarwal:2012} and in the present work, respectively.

\begin{table}[htbp!]\centering
    \caption{\normalsize{Transition energy $E$ and lifetime $\tau$ of the 1s$^2$--1s2p~$^1 \! P_1$ transitions in the He-like Ca ion.}}
    \begin{tabular}{llcc}
    \hline\hline
    \multicolumn{1}{c}{$E$ [eV]\phantom{ene}} & 
    \multicolumn{1}{c}{$\tau$ [fs]\phantom{en}} & year & reference \\
    \hline 
    \phantom{390} 
    ---       & 6.06  & 1977 & \cite{LinJohnsonDalgarno:1977} \\
    3902.3676    & ---   & 1988 & \cite{Drake:1988} \\
    3902.3775(4) & ---   & 2005 & \cite{Artemyev:2005} \\
    3902.2570    & 5.946 & 2012 & \cite{Aggarwal:2012} \\
    3902.2551    & ---   & 2021 & \cite{NIST-ASD} \\
    3902.3351    & 6.09  & 2021 & this work \\
    \hline\hline
\end{tabular}
\label{table.He-like-Ca}
\end{table}

\subsection{Li-like Ca}
\label{section.Li-like-Ca}

Similarly as in the case of He-like Ca, for Li-like Ca ion the transition energies and lifetimes were calculated using {\sf GRASP} package.  The results for four different excitations from the ground {1s$^2$2s} state are presented in Table~\ref{table.Li-like-Ca} and compared with data available in the literature.

\begin{table*}[!htbp]\centering
    \caption{\normalsize{Transition energies $E$ and lifetimes $\tau$ of the 2s--2p$_{1/2}$, 2s--2p$_{3/2}$, 2s--3p$_{1/2}$, 2s--3p$_{3/2}$ transitions in the Li-like Ca ion.}}
    {\small 
    \begin{tabular}{ll|ll|ll|ll|clc}
        \hline
        \hline 
        \multicolumn{2}{c|}{2s--2p$_{1/2}$} & 
        \multicolumn{2}{c|}{2s--2p$_{3/2}$} & 
        \multicolumn{2}{c|}{2s--3p$_{1/2}$} & 
        \multicolumn{2}{c}{2s--3p$_{3/2}$} \\
        \hline
        \phantom{3}$E$~[eV] & $\tau$~[ns] & 
        \phantom{3}$E$~[eV] & $\tau$~[ns] & 
        \phantom{3}$E$~[eV] & $\tau$~[ps] & 
        \phantom{3}$E$~[eV] & $\tau$~[ps] & year & method & ref. \\
        \hline
        &  0.76752 &        & 0.51258  &       &       &       &       & 1991 & Coul-App HS-core & \cite{TheodosiouCurtisEl-Mekki:1991} \\
        \hline
        35.963  & 0.7680    &  41.029 & 0.5123   &      &          &       &            & 1996 & 3-rd order MBPT & {\cite{JohnsonLiuSapirstein:1996}} \\
        \hline
        &          &       &      &  663    &  0.4167  &  663      & 0.4274 & 2002 & R-matrix Breit-Pauli & \cite{Nahar:2002} \\ 
        \hline
        35.962{(1)}  &           & 41.024{(1)} &        &          &          &       &            & 2011 & S-matrix Kohn-Sham &  {\cite{SapirsteinCheng:2011}} \\
        \hline
        &          &       &      &  661.7643 & 0.4282 &  663.2660 & 0.4367  & 2014 & MCDF VP SE & \cite{DengJiangZhang:2014} \\
        \hline
        35.96119{(7\rlap{3)}} &       & 41.02497{(7\rlap{8)}} &     &       &          &       &            & 2018 & RCI QED NucPol & {\cite{YerokhinSurzhykov:2018b}} \\ 
        \hline
        \multicolumn{1}{l}{35.9625 } & & 
        \multicolumn{1}{l}{41.0286 } & & 
        \multicolumn{1}{l}{661.8896 } & & 
        \multicolumn{1}{l}{663.3403 }  & & 2021 & NIST ASD & \multicolumn{1}{c}{\cite{NIST-ASD}} \\
        \hline
        35.959  & 0.767   &   41.027  & 0.512  &  661.776 & 0.428   &  663.278  & 0.436   & 2021 &  RCI VP SE & this work \\ 
        \hline
        \multicolumn{10}{c}{experiment} \\ 
        \hline
        35.962{(2)}  &          &        &         &       &       &       &       & 1985 &  & \cite{SugarCorliss:1985} \\
        \hline
        &      & 41.029{(2)} &         &       &       &       &       & 1983 &  & \cite{Edlen:1983} \\
        \hline
        \hline
        \vspace{0.1cm}
    \end{tabular}
    }
    \label{table.Li-like-Ca}
\end{table*}

\subsection{Bright future of partially stripped ions in the Gamma Factory}
\label{section.futurePSI}

The Li-like Pb ions will be accelerated and irradiated in the GF PoP experiment at the SPS \cite{GF-PoP-LoI:2019}.  During the experiment, various aspects of the project, including the efficiency of the PSI excitation, will be studied.  This will be the next step in the project which may open means for the GF implementation at the Large Hadron Collider (LHC).  For that experiment, the H-like Pb ion with much larger transition energy is considered \cite{Bessonov:1995eq}.  The transition becomes accessible for existing light sources, such as the Free Electron Laser (FEL), in conjunction with a higher value of the Lorentz factor $\gamma_L$ of the PSI bunches.
The Li-like Ca ions are suggested for the high-luminosity version of the LHC with laser-cooled isoscalar ion beams \cite{Krasny:2020wgx}, while the He-like Ca ions can be used for the potential future production of the radioactive-ion beams at the GF \cite{Nichita:2021iwa}.  In the following section we analyze the interaction of the laser light with the PSI bunch circulating in these two accelerators, investigating different scenarios of the process. 
\section{Exciting ultra-relativistic ions with light}
\label{sec:OptPump}

In high-energy physics, a cross section is often used to describe a scattering process. While in optics the cross section is also used, its application implicitly assumes that the process is investigated in the steady state, when dynamic equilibrium between different processes (e.g., excitation and relaxation) is reached.  Prior to the steady state, however, the system experiences a transient period during which it may undergo significant changes.  The dynamics of this transient evolution depends on many parameters, including incident-light intensity, strength of an atomic transition,
light detuning or excited-state relaxation.  Thereby, over time comparable with the excited-state lifetime $\tau_e$ the population of the excited state, which determines the intensity of the fluorescence from the ions (see below), may continuously increase, eventually reaching its steady-state value, but it may also experience oscillations before finally leveling up at a specific value.  As the frequency and amplitude of these, so-called, Rabi oscillations depend on parameters of incident light, studies of the dynamics of PSI excitation at times shorter than the excited-state lifetime become an important aspect of the GF.

Below, we analyze, both theoretically and numerically, the problem of optical excitation of an energy-dispersed ion bunch by a pulse of light.  By investigating the interaction of a resonant light pulse with a generic closed two-level system, i.e., with a system where levels other than these directly coupled by light are ignored, we analyze the situation which, to the first order, reproduces the GF set-up.  

\subsection{Theoretical model}

We consider an excitation of a two-level atom with semi-resonant light, $\Delta\omega\ll\omega$, where $\omega$ is the light frequency and $\Delta\omega$ is its detuning from the optical transition.  Since the two-level system is considered, there are no dark states, which could prevent the ions from further excitation.  In this system, the atoms are characterized with the excited-state relaxation rate $\gamma_e$, and we also assume that the ground-state lifetime is infinite, $\gamma_g=0$. Finally, the interaction is considered in the rotating-wave approximation, when interaction with only a resonant components of the light field ($\omega\approx\omega_0$) is considered, while the effect of the second (conjugate) frequency component of light, $-\omega$, is neglected.

In order to determine scattering of photons by the atoms, the time-dependent expectation value of the spontaneous-emission operator $\mathcal{F}$ needs to be calculated.  Herein, we calculate the value using the density-matrix formalism
\begin{equation}
	\langle\mathcal{F}\rangle=\textrm{Tr}(\rho\mathcal{F}),
	\label{eq:expF}
\end{equation}
where $\rho$ is the density matrix of the atoms \cite{AuzinshBook2010}.  Evolution of the density matrix can be described using the Liouville equation
\begin{equation}
	\dot{\rho}=\frac{i}{\hbar}[H,\rho]-\frac{1}{2}\left\{\Gamma,\rho\right\},
	\label{eq:Liouv}
\end{equation}
where $H$ is the Hamiltonian of the system, containing the contribution from the Hamiltonian of the unperturbed atoms $H_0$ and the operator $V$, describing their interaction with light. The operator $\Gamma$ describes the relaxation in the system, in particular the relaxation of the excited state due to spontaneous emission.  It can be shown \cite{AuzinshBook2010} that, in the case of a two-level system, the matrix elements of the fluorescence operator are given by
\begin{equation}
	    F^{e}_{g}=\frac{4}{3}\,\frac{\omega_{0}^3}{\hbar c^3}\;\vec{d}_{ge}\cdot\vec{d}_{eg},
	\label{eq:Fge}
\end{equation}
where $\vec{d}_{eg}$ is the electric dipole moment between the ground state $g$ and the excited state $e$.  Because the electric dipole moment is an odd operator, the only nonzero elements of the fluorescence operator $\mathcal{F}$ are at a diagonal.  Moreover, since the fluorescence arises exclusively due to spontaneous emission, and the ground state is relaxation free, the fluorescence of the atoms is proportional to the excited-state population $\rho_{ee}$.  Hence the time-dependent fluorescence operator expectation value is given by
\begin{equation}
	\langle\mathcal{F}\rangle= \frac{3\gamma_e N_{\rm PSI}}{\hbar c^3}\,\rho_{ee},
	\label{eq:FluorescenceFinal}
\end{equation}
where $N_{\rm PSI}$ is the number of partially stripped ions.

Since the only dynamic parameter in Equation~\ref{eq:FluorescenceFinal} is the excited-state population $\rho_{ee}$, henceforth we investigate evolution of the population.  Moreover, normalization of the population, i.e., $\rho_{gg}+\rho_{ee}=1$, where $\rho_{gg}$ is the ground-state population, allows to relate the population $\rho_{ee}$ with the probability of the excited-state occupation.  This provides an intuitive insight into the efficiency of PSI excitation/fluorescence; the higher the excited-state population, the more intense fluorescence from the ions.

The problem of the interaction of classical light with the two-level atom using the Liouville equation is considered in many textbooks (see, for example, Ref.~\cite{AuzinshBook2010}).  The evolution of the density matrix elements are given by
\begin{eqnarray}
        \dot{\rho}_{eg}&=&\left(i\Delta\omega-\frac{\gamma_e}{2}\right)\rho_{eg}+\frac{i\Omega_R}{2}(\rho_{gg}-\rho_{ee}),\label{eq:LiouvilleEquationGeneralFirst}\\ 
        \dot{\rho}_{ge}&=&-\left(i\Delta\omega+\frac{\gamma_e}{2}\right)\rho_{ge}-\frac{i\Omega_R}{2}(\rho_{gg}-\rho_{ee}),\\ 
        \dot{\rho}_{ee}&=&\frac{i\Omega_R}{2}(\rho_{ge}-\rho_{eg})-\gamma_e\rho_{ee}, 
        \label{eq:LiouvilleEquationGeneralLast}
\end{eqnarray}
where $\rho_{eg}$ is the envelope of optical coherence (an amplitude of the superposition between the ground state $g$ and excited state $e$) and 
\begin{equation}
\Omega_R=c\,\sqrt{\frac{6\pi\gamma_e I}{\hbar\omega_0^3}}
\label{eq:RabiFreq}
\end{equation}
is the Rabi frequency, characterizing the coupling strength between light and ions, with $I$ being light intensity.  Solving this set of equations allows one to determine the excited-state population, and hence the number of fluorescence photons.

\subsection{Scattering at steady state}
\label{subsec:SteadyState}

Let us first consider the stationary situation when equilibrium between various processes is achieved, i.e., the steady-state situation.  In such a regime, the left-hand sides of Equations~\ref{eq:LiouvilleEquationGeneralFirst}--\ref{eq:LiouvilleEquationGeneralLast} are equal to zero, $\dot{\rho}=0$, which, through algebraic manipulations of Equations \ref{eq:LiouvilleEquationGeneralLast} allows us to calculate the excited-state population
\begin{equation}
\rho_{ee}=\frac{\Omega_R^2/4}{\Delta\omega^2+\gamma_e^2/4+\Omega_R^2/2}=\frac{\kappa_1/2}{1+4\Delta\tilde{\omega}^2+\kappa_1},
    \label{eq:PopulationSteadyState}
\end{equation}
where $\Delta\tilde{\omega}=\Delta\omega/\gamma_e$ is the normalized detuning and $\kappa_1=2\Omega_R^2/\gamma_e^2$ is the saturation parameter, relating the strength of the light--atom coupling (given by the Rabi frequency $\Omega_R$) to the system's relaxation $\gamma_e$.   In particular, Equation~\ref{eq:PopulationSteadyState} shows that the excited-state population, and hence fluorescence, depends on the light intensity and detuning.  

Comparison of Equation~\ref{eq:PopulationSteadyState} with the classical absorption cross section \cite{Hulst2012Light},
\begin{equation}
    \sigma=\frac{\sigma_0}{1+4\Delta\omega^2/\gamma_e^2+2\Omega_R^2/\gamma_e^2}=\frac{\sigma_0}{1+4\Delta\omega^2/\gamma_t^2}\,,
    \label{eq:AbsorptionCrossSection}
\end{equation}
where $\sigma_0$ is the resonant absorption cross section and $\gamma_t$ is the transition linewidth, reveals similarity between the classical and quantum approach.  In particular, both approaches show that the further the light is detuned from the optical transition, the less efficient the excitation is.  In both approaches scattering also depends on the transition linewidth $\gamma_t$ (full width at half maximum -- FWHM), which in the classical approach is light-intensity independent and is determined by the excited-state relaxation rate $\gamma_e$, $\gamma_t=\gamma_e$.  The difference between the classical and quantum mechanical approach arises at higher light intensities.  In such a case, the quantum-mechanical approach incorporates the saturation effect, which modifies the effective linewidth of the transition, $\gamma_t=\gamma_e \Delta\omega/\sqrt{\Delta\omega^2+\Omega_R^2/2}$ and causes leveling up the efficiency of the excitation at 1/2 for $\Omega_R^2\gg \Delta\omega+\gamma_e^2/4$.  The saturation effect stems from the finite number of atoms that can be excited by light and finite time the excited atom needs to emit the photon (as discussed above, the excited atoms emits a photon with the characteristic time $\tau_e$).  This means that at some point further increase of the incident light intensity does not result in the increase of the number of absorbed/scattered photons as all atoms are already involved into scattering.  The intensity at which the saturation occurs depends on the light detuning, i.e., saturating the transition with off-resonance light requires more intense incident light than it is the case for on-resonance light.
\begin{figure}[!htbp]
    \centering
    \includegraphics[width=0.95\linewidth]{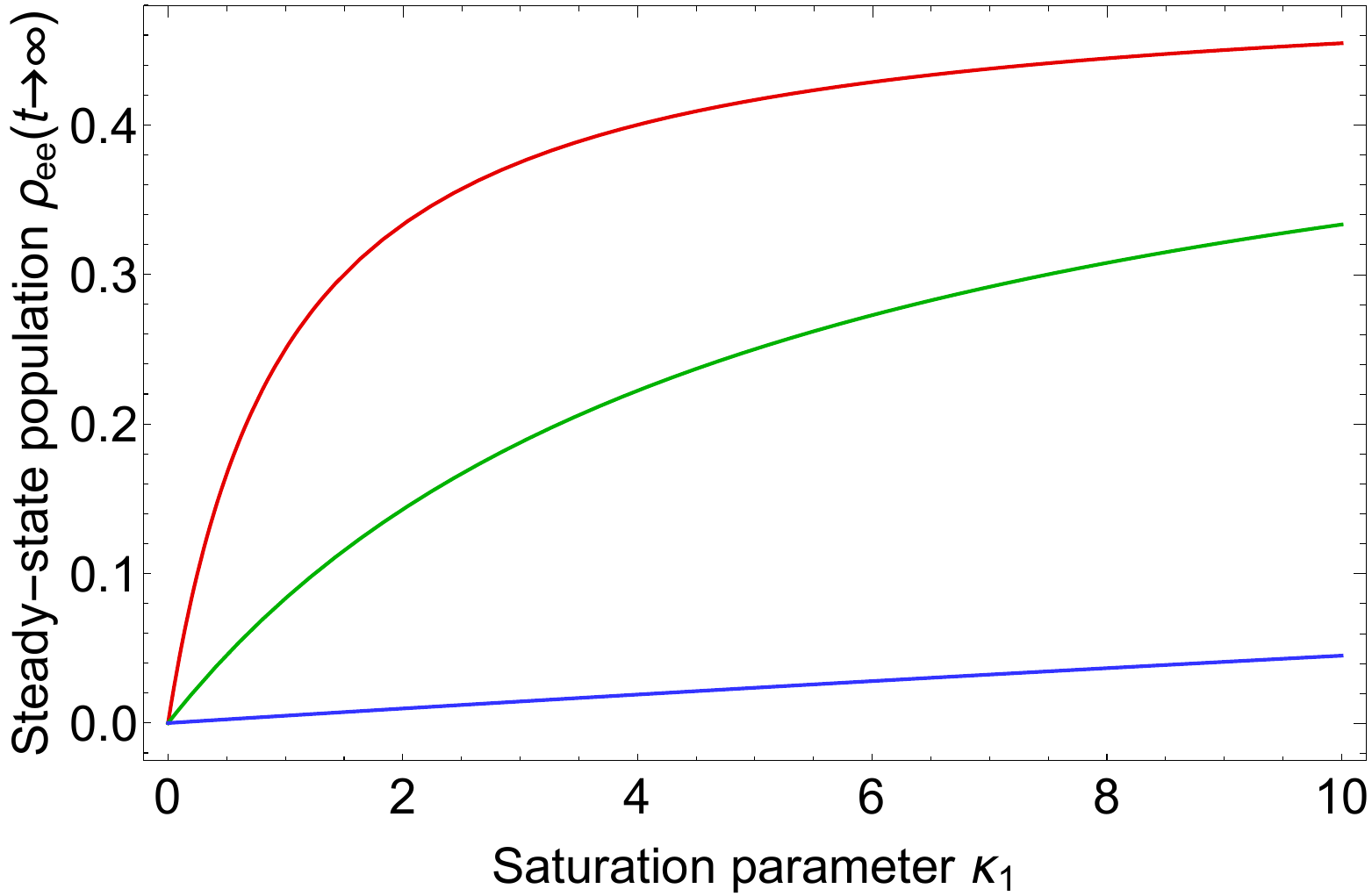}
    \caption{The steady-state excited-state population $\rho_{ee}$, determining the fluorescence of the PSI illuminated with light, versus the saturation parameter $\kappa_1$ for three different normalized detunings: $\Delta\tilde\omega=0$ (red), $\Delta\tilde\omega=1$ (green) and $\Delta\tilde\omega=-5$ (blue).}
    \label{fig:FluorescenceSteadyState}
\end{figure}

The discussion presented above is depicted with the quantum-mechanical results shown in Figure~\ref{fig:FluorescenceSteadyState}, where the excited-state population is plotted against the saturation parameter $\kappa_1$ for three different detunings.  As expected, increasing the light power (saturation parameters) improves the excited-state population and hence the intensity of the scattered light.  While initially the process linearly depends on incident-light intensity, at higher intensity it begins to saturate.  The results also show that saturating the transition with detuned light requires higher light intensity.  This dependence also indicates that saturating moving ions, whose transition frequencies in the laboratory frame (LF) are modified due to the Doppler effect, with a CW light, is more challenging than in the case of motionless ions.

The remaining question is how fast the system reaches the steady state, which can be rephrased into: when the classical approach still adequately describes the atom--light interaction (even if saturation is somehow taken into account in the classical approach).  From Equations~\ref{eq:LiouvilleEquationGeneralFirst}--\ref{eq:LiouvilleEquationGeneralLast} one can generally conclude that the system reaches the steady state for times on the order of $\tau_e$  (more careful analysis reveals that steady-state population is reached at $t\approx 5\tau_e$).  This reveals a fundamental role of spontaneous emission, which acts as a dephasing mechanism for the Rabi oscillations;  initially, all the ions oscillate in phase, but every time spontaneous emission occurs the phase of the ion is randomized. As a result, after the time comparable with the excited-state lifetime, the Rabi phase of majority of the ions is reset and the system reaches the ``incoherent" steady state with a given distribution between the ground and excited state population. Simultaneously, if the interaction time is shorter than $\tau_e$, the classical approach using the cross section does not work and dynamics of the system needs to be evaluated using the quantum-mechanical formalism.  This evolution is discussed in the following section.

\subsection{Dynamics of ion excitation\label{Sec:DynamicsPumping}}

The ultrarelativistic nature of the GF ions results in a significant difference in the flow of time in the ion-rest and laboratory frames.  As a result, the excited-state lifetime in the ion-rest frame (IRF) $\tau_e$ corresponds to the LF excitation time $\tau_e^{\rm LF}$ via
\begin{equation}
    \tau_e^{\rm LF}=\gamma_L\,\tau_e.
\end{equation}
As a consequence, the average path an excited ion propagates in the LF after the excitation is
\begin{equation}
    l^{\rm LF}=c\gamma_L\,\tau_e.
\end{equation}

\begin{figure}[!htbp]
    \centering
    \includegraphics[width=0.95\linewidth]{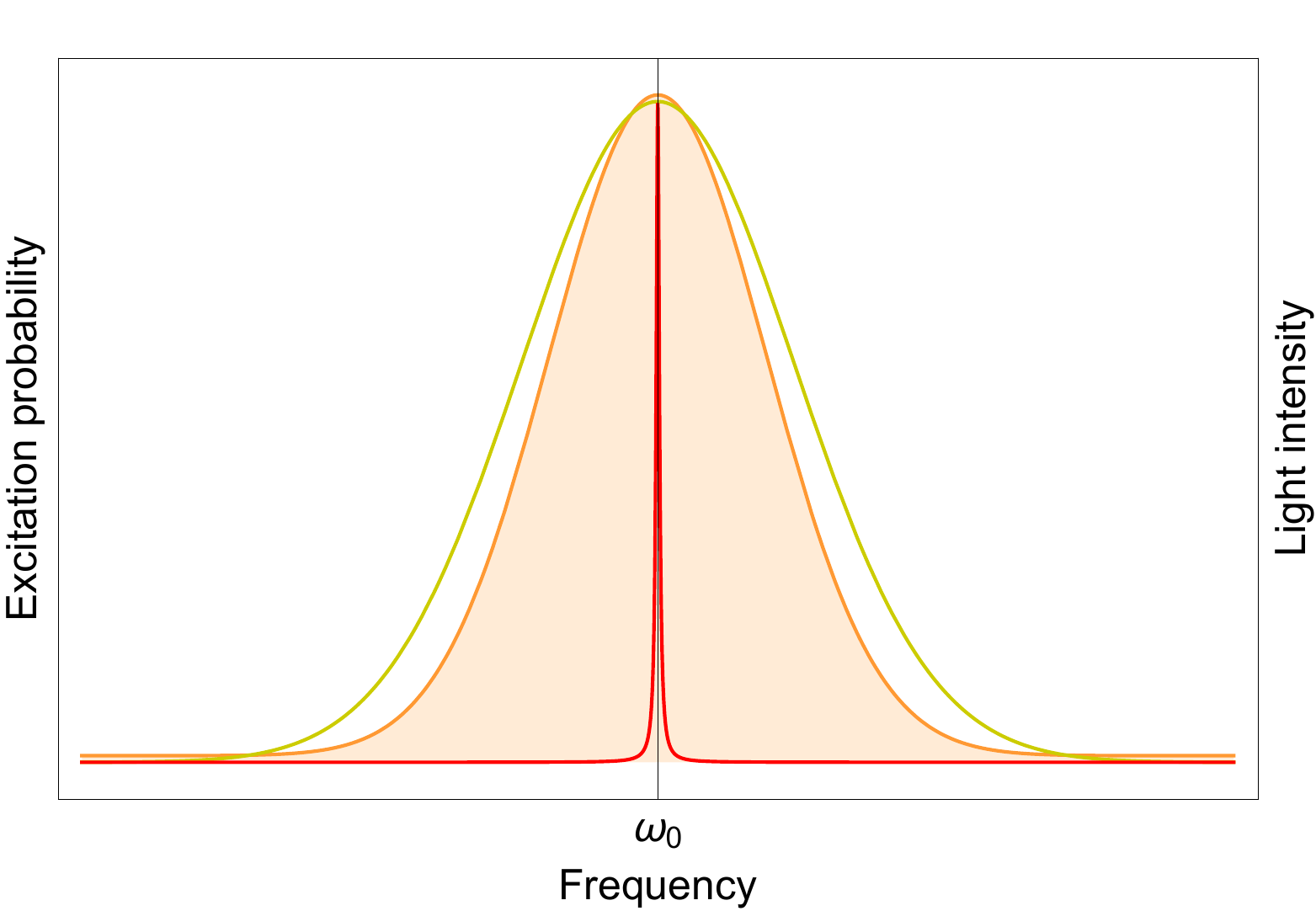}
    \caption{Schematics of spectral characteristics of the system. The red line shows the probability of excitation of the motionless PSI. In a weak-light regime, the probability is given by the Lorentz function whose full width at half maximum (FWHM) is determined by the excited-state relaxation rate $\gamma_e$. The middle orange line corresponds to the spectral profile of the pulse used for the excitation, where shading indicates the frequency range that can be used for the ion excitation. The broadest profile corresponds to the transition line inhomogeneously broadened due to the Doppler effect.}
    \label{fig:SchematicExcitation}
\end{figure}

The ions intended to be used in the GF will be excited at a relatively narrow transition, $\omega_0\gg\gamma_e$.  However, due to the ion energy dispersion 
$\Delta \mathcal{E}/\mathcal{E}=\Delta\gamma_L/\gamma_L$, the transition is inhomogeneously broadened (the Doppler effect).  In the PoP experiment \cite{GF-PoP-LoI:2019}, the Doppler broadening of the transition is 2--4 orders of magnitude larger than its natural width $\gamma_e$, as schematically depicted in Figure~\ref{fig:SchematicExcitation}.  As shown with Equation~\ref{eq:PopulationSteadyState}, to saturate a transition with detuned light requires higher intensities. In fact, to do so for the light detuned by $\Delta\omega$ the intensity should be roughly $4\Delta\omega^2/\gamma_e^2$ times higher. In turn, to saturate the whole Doppler-broadened transition with CW light, the light intensity in the PoP experiment would have to be increased by 4--8 orders of magnitude. On the one hand, this may be difficult, if possible at all, but more importantly, due to the finite interaction region, the ions do not experience the CW light but rather a light pulse of the Fourier-broadened spectrum.  In fact, we exploit this effect to facilitate the interaction and more efficiently excite the ions.  Specifically, we aim at generating pulses, which spectral width coincides with the Doppler-broadened transition of the ions on the bunch
\begin{equation}
    t_p^{\rm IRF}=\frac{2\gamma_L}{\sigma^{\rm LF}_\omega},
\end{equation}
where $\sigma^{\rm LF}_\omega$ is the root-mean-square (rms) pulse width in the lab frame.

The spectral broadening of the pulse significantly beyond the transition natural linewidth has an important consequence.  Specifically, it allows to neglect the relaxation of the ions during the interaction, which significantly simplifies the theoretical description.  In fact, the interaction of light with the relaxation-free two-level atom is a textbook example (see, for example, Ref.\ \cite{Letohov1977Nonlinear}), demonstrating oscillations of the excited-state population $\rho_{ee}(t)$ at the Rabi frequency $\Omega_R$ (the Rabi oscillations).  While dynamics of this coherent, i.e., uninterrupted by spontaneous emission, evolution is harder to determine in the case of pulsed excitation, where the Rabi frequency varies over time, we can generally state that the excited-state population at the time moment $t_1$ is given by $\sin^2\left(\int_0^{t_1}\Omega_R(t)dt\right)$, where the expression under the sine function is the total Rabi phase of the oscillation.  In fact, the dependence may enable mimicking the short-pulse (dynamic) regime even in Monte-Carlo simulations.  This would be the case when one is not interested in describing the whole evolution of the system during the pulse but rather aims at the state of the atoms after transition of the pulse.

From the perspective of the GF, the last consequence of the short length of the pulse is the absence of ion--ion interactions (e.g., collisions) during the light pulse.  Thereby, the PSI in different velocity classes can be treated independently and the problem can be further simplified and the excited-state population of the whole bunch is simply a weighted average over the ions' velocity distribution.  

\begin{figure*}[!htbp]
    \centering
    \includegraphics[width=1\linewidth]{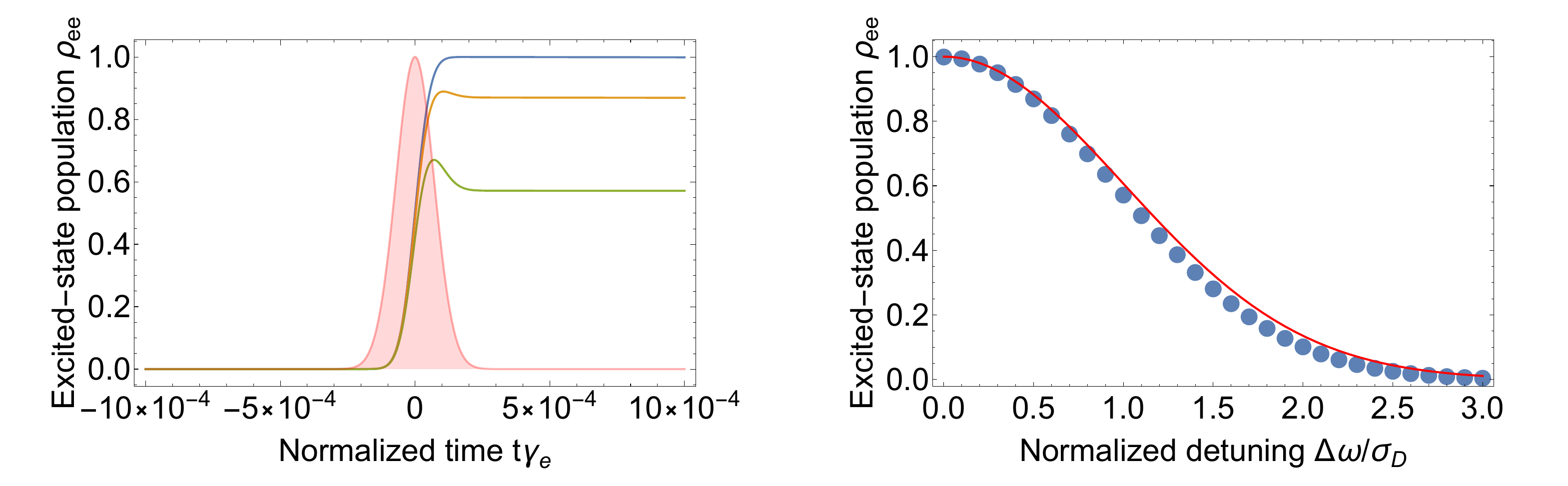}
    \caption{(Left) The excited-state population of the PSI interacting with the Gaussian light pulse (red trace) of a spectral width coinciding with the atoms' Doppler profile, $\gamma_p^{\rm IRF}=\sigma_\omega^{\rm IRF}$.  Different traces corresponds to different detunings of the light central frequency $\omega$ from the Doppler-shifted resonance frequency: $\Delta\omega/\sigma_\omega^{\rm IRF}=0$ (blue), $\Delta\omega/\sigma_\omega^{\rm IRF}=0.5$ (yellow) and $\Delta\omega/\sigma_\omega^{\rm IRF}=1$ (green).  (Right) The population of the excited states after the pulse (blue dots) along with the number of the PSI in a specific velocity class (red line) versus the normalized detuning.  The results indicate that the PSI-distribution averaged population of the exited state is 70\%. The simulations were performed for the pulse spectrally covering the whole inhomogeneously broadened spectral line and the amplitude of the pulse $\Omega_R^0\approx18000\,\gamma_e$.}
    \label{fig:RabiOscillationsPulsed}
\end{figure*}

The final stage of our discussion concerns the efficiency of the excitation of atoms with different detunings.  In the case of the interaction with the light pulse spectrally coinciding with the Doppler width, the different velocity classes are resonantly excited by appropriate spectral components of the light.  This alleviates the demanding requirement for the light intensity, which may lead to the problems with photoionization or multiphoton excitation.  Moreover, this also provides a better control over the efficiency of the interaction. As shown in Figure~\ref{fig:RabiOscillationsPulsed}, the interaction with a pulse, whose central frequency $\omega$ is detuned from the Doppler-shifted resonant frequency by the rms Doppler width, $\Delta\omega=\sigma_\omega^{\rm LF}$ reduces the excitation efficiency by half. In the case of spectrally narrow light, this case would correspond to very small excitation, unless extremely intense light is used. 

Under the assumption of a short light pulse, the efficiency of the excitation of the whole PSI bunch can be calculated by averaging over the distribution due to the energy dispersion. When the pulse amplitude is chosen in such a way that in-resonance ions undergo half of the Rabi cycle -- the whole population is transferred into the excited state -- and the pulse width coincides with the ions Doppler broadening, $\gamma_p^{\rm IRF}=\sigma_\omega^{\rm IRF}$, the efficiency of excitation of complete ion bunch would reach 70\%.

\subsection{Parameters of envisioned experiments}

Let us now discuss the excitation of PSI in the main three scenarios considered for the GF: lithium-like lead, planned to be used in the PoP experiment in the SPS \cite{GF-PoP-LoI:2019}, hydrogen-like lead, envisioned for the LHC experiment \cite{Bessonov:1995hd}, and lithium-like calcium, considered for optical cooling of accelerator beams \cite{Krasny:2020wgx}.  
The case of the helium-like calcium, discussed in Section~\ref{section.IonChoice},
is not considered here but will be presented in our future studies on other possible applications of the GF.
Table~\ref{table:OpticalPumpingParameters} summarizes the physical parameters based on our assumptions, data from the literature, and results of calculations performed in this work.

\begin{table*}[!htbp]
    \caption{The optical parameters for the planned GF experiments. 
    The relative Gaussian energy spread of $2\times 10^{-4}$ for both the PSI bunch and the laser pulse is assumed 
    in all cases.}
    \centering
    \begin{threeparttable}
    \begin{tabular}{l|c|c|c}
    \hline\hline
    Parameter name & 
    Li-like ${}^{208}_{\phantom{0}82}$Pb$^{79+}$ & 
    Li-like ${}^{40}_{20}$Ca$^{17+}$ & 
    H-like ${}^{208}_{\phantom{0}82}$Pb$^{81+}$ \\
         \hline
         {Electronic transition} & $2s\rightarrow 2p_{1/2}$ &  $2s\rightarrow 3p_{1/2}$ & $1s\rightarrow 2p_{1/2}$ \\
         {Transition energy $\omega_0$ [eV]} & $230.823\,(47)(4) $ \cite{YerokhinSurzhykov:2018a,YerokhinSurzhykov:2018b} & $661.89$ \cite{NIST-ASD} & $75\,280.83\,(26)$ \cite{YerokhinShabaev:2015} \\
         {Excited-state lifetime $\tau_e$ [ps]} & $76.6$ \cite{JohnsonLiuSapirstein:1996} & $0.43$\tnote{(a)} & $3.4\times 10^{-5}\,$\tnote{(a)}  \\
         Excited-state relaxation rate $\gamma_e$ [s$^{-1}$] & $1.3\times 10^{10}$ & $2.3\times 10^{12}$& $3.0\times 10^{16}$ \\
         rms Doppler width $\sigma_\omega$ [s$^{-1}$] & $7.0\times 10^{13}$ & $2.0\times 10^{14}$ & $2.3\times 10^{16}$\\ \hline
         Pulse energy [mJ] & 0.2 and 5.0 & 0.35 and 2.0 & $5.0\,$\tnote{(b)} \\
         LF radiation energy [eV] & 1.2 & 1.6 & 12.6\\
         rms LF pulse length $\tau_p^{\rm LF}$ [ps] & $2.8$ & $2.0$& $500$  \\
         rms\ transverse pulse size $\sigma^p_x=\sigma^p_y$ [m] & $6.5\times 10^{-4}$ & $5.6\times 10^{-4}$ & $2.5\times 10^{-5}$  \\ \hline
         Number of ions per bunch $N_b$ & $9\times 10^7$ & $4\times 10^9$& $9.4\times 10^7$\\
Lorentz relativistic factor of ion $\gamma_L$ & $96.3$ & $205.62$ & $2989$ \\
         rms\ transverse ion beam size [m] & $\sigma_x = 10.5 \times 10^{-4}$ & $\sigma_x=8.0 \times 10^{-4}$ & $\sigma_x = 3.9\times 10^{-5}$ \\
          & $\sigma_y=8.3 \times 10^{-4}$ &  $\sigma_y=5.7 \times 10^{-4}$ & $\sigma_y = 3.9\times 10^{-5}$ \\
         rms\ ion bunch length $\sigma_z$ [m] & $0.06386$ & $0.10$& $0.15$  \\ \hline
         rms IRF pulse spectral width $\gamma_p^{\rm IRF}$ [s$^{-1}$] & $6.9\times 10^{13}$ & $1.4\times 10^{14}$ & $8.4\times 10^{12}$ \\
         Spatio-temporal IRF Rabi amplitude $\Omega_0^{\rm IRF}$ [s$^{-1}$] & $6.2\times 10^{13}$ and $3.1\times 10^{14}$ & $6.7\times 10^{14}$ and $1.6\times 10^{15}$ & $4.8\times 10^{15}$\\
         Characteristic LF excitation distance $x^{\rm LF}$ [m]  & $2.2$ & $0.027$ & $3\times 10^{-5}$ \\ \hline
         Number of photons emitted from the bunch  & $2.0\times 10^6$ and $8.6\times 10^{6}$ & $5.0\times 10^8$ and $5.7\times 10^8$ &  $3.8\times 10^{8}$\\
         \hline\hline
    \end{tabular}
    \begin{tablenotes}\footnotesize
      \item[(a)] This work.
      \item[(b)] Power inaccessible for the current light sources at wavelength of $\approx 100$\,nm, but anticipated in the future.
    \end{tablenotes}
    \end{threeparttable}
    \label{table:OpticalPumpingParameters}
\end{table*}

The ion bunch is characterized with a three-dimensional Gaussian function with the rms widths $\sigma_x$, $\sigma_y$, and $\sigma_z$ in all three directions with the dominant width along $z$ (a cigar-like shape).  In the interaction region, the light-pulse-intensity profile is given by a symmetric Guassian function of a transverse width of $\sigma^p_x=\sigma^p_y$.  In general, the transverse sizes of light pulse and bunch are not matched, which affects the efficiency of the interaction (a number of light accessible ions limited by a geometrical factor).  The efficiency may be further reduced by a non-zero angle between laser and ion beam directions, i.e., imperfect anti-collinear alignment of the two beams.  The excitation of PSI is induced by light pulse, which central frequency $\omega$ is tuned to the center of the Doppler broadened transition $\omega_0$, $\omega=\omega_0$.

All the ions considered in this work are characterized with the energy-level structure of a total angular momentum of 1/2 in the ground state and a total angular momentum of 1/2 in the excited state.  Despite the fact that there are two magnetic sublevels in either of the states, as well as specific selection rules associated with the excitation for a given light polarization, it can be shown that such a system may be effectively reduced to the two-level system discussed above.

In the PoP experiment at the SPS, lithium-like lead ions ($^{208}_{\phantom{0}82}{\rm Pb}^{79+}$) will be used.  With the Lorentz factor $\gamma_L=96.3$ and a transition energy between the two lowest electronic levels of $\sim230\,$eV, one can show that the transition can be induced with a Ti:sapphire laser, emitting infra-red radiation at $1035\,$nm.  Motivated by the technical limitations, we have chosen the pulse of the LF length of 2.8\,ps (rms width), i.e., the pulse spectral width covers 70\% of the bunch Doppler width.  While the interaction efficiency depends on temporal and spectral parameters of the experiment, it is difficult to {\em a priori} determine the parameters of the pulse.  Therefore, we consider two pulse energies: 0.2\,mJ and 5\,mJ.  The first pulse transfers about 35\% of the zero-detuning atoms residing in the center of the bunch, $x=y=0$.  Despite such a high efficiency of excitation in the center, accounting for the beam and bunch profiles significantly modifies the excitation, so that the overall efficiency of the bunch excitation reaches just 2.2\%.  Increasing the pulse energy by a factor of 25, to 5\,mJ, results in roughly one full Rabi cycle experienced by the ions in the bunch center that are tuned to the resonance.  As very few ions of the group are being excited, this may suggest that the overall efficiency of excitation is smaller than in the previous case.  However, because of the experimental geometry (non-zero angle between laser and ion beam directions), increasing the light power results in excitation of the ions that remained in the ground state. In turn, the overall efficiency of the ion excitation rises from 2.2\% to 9.6\%.  The correspond single-bunch scattered-photon number rises from $2.0\times 10^6$ in the first case to $8.9\times 10^6$ in the second case.  The presented results suggest that increasing the pulse energy increases the excitation efficiency.  Indeed, for the energy range between 0 and 2.5\,mJ, the efficiency monotonically increases in nonlinear manner, reaching its maximum of 10.7\% at 2.5\,mJ.  For higher energies, the efficiency drops, which is an indication of complex dynamics of the system, yet the dependence is much weaker (from 2.5\,mJ to 5\,mJ the redaction is from 10.7\% to 9.6\%).  Independently from the actual excitation efficiency, the photons will be emitted in the forward cone within several nanoseconds over a distance of a few meters.

A similar situation is encountered in the case of the lithium-like calcium ion ($^{40}_{20}$Ca$^{17+}$), i.e., the Doppler-broadened transition is several orders of magnitude larger than the natural linewidth.  It should be noted, however, that the difference between linewidths is smaller than in the previous case, which manifests via a larger contribution of the relaxation to the system evolution.  With a Lorentz factor of 205, the PSI may be excited with the light pulse of the carrier wavelength of $771\,$nm, and the pulse length of $2\,$ps allows to spectrally cover 70\% of the Doppler-broadened transition.  Performing the optimization of the pulse energy for given pulse parameters, we have found that below 1\,mJ, the maximum excitation intensity of 12.5\% is achieved for the pulse energy of 0.35\,mJ (about 14.2\% of excitations can be achieved for the pulse energy of 2\,mJ).  This corresponds to about $5.0\times 10^{8}$ ($5.7\times 10^{8}$) photons emitted from each PSI bunch within about $10\,$ps over a distance of about $3\,$mm in the LF.

A different situation is encountered in the case of hydrogen-like lead ions ($^{209}_{\phantom{0}82}{\rm Pb}^{81+}$).  As shown in Table~\ref{table:OpticalPumpingParameters}, the natural linewidth of the transition is comparable with the Doppler width.  Thus, it can be assumed that even with a spectrally narrow pulse all the ions are excited with a comparable efficiency. Specifically, the efficiency drops by a factor of $2$ for the detuning $\gamma_e/2$. Another important difference is the light-induced evolution of the excited-state population, determining the fluorescence of the ions.  In contrast to the previous cases, the pulse is orders of magnitude longer than the excited-state lifetime, thus the system reaches the equilibrium during each instant of the optical pulse. Thereby, the classical cross-section approach can be used in the considered case.  Moreover, the difference in the excited-state lifetime and pulse length suggests that during the light pulse the ions may undergo multiple excitation--emission cycles.  This significantly enhances the photon emission from the ions, even though the efficiency of the excitation is low.  Our calculations show that despite the very low excitation efficiency (at most 0.4\%) with the LF light of the carrier wavelength of 98.46 nm, the pulse energy of 5\,mJ and the pulse rms length of 500\,ps, i.e., the parameters that are not accessible experimentally at the moment but are foreseen in the future, the ions can be excited many times during the pulse.  In turn, under the experimental conditions, each ion can be excited 4 times on average, corresponding to $3.8\times 10^8$ emitted photons for each bunch.  Due to the low saturation, the number of emitted photons would scale linearly with both the length and energy of the pulse, revealing the room for further improvements.

An important question in our consideration is a potential presence of additional energy levels that may trap the ions, making them inaccessible to light. Such levels of energies lower than the energies of the corresponding excited states do not exist in the case of $^{208}_{\phantom{0}82}{\rm Pb}^{79+}$ and $^{208}_{\phantom{0}82}{\rm Pb}^{81+}$ ions. Albeit there is such a level in the $^{40}_{20}{\rm Ca}^{17+}$ ion, in the considered scenario of interaction with a light pulse that is several orders of magnitude shorter than the excited-state lifetime, the spontaneous decay can be neglected, so that this level may not contribute to the dynamics of the interaction. Principally, the problem might be with higher-energy levels present in the ions. However, for the laser-pulse intensities considered in this paper, the probability of multi-photon absorption to these levels is low, so that their effect can also be neglected. In turn, the two-level model well describes the systems under consideration. 
\section{Simulations of optical excitation}
\label{sec:MCsim}

As discussed in Subsection~\ref{subsec:SteadyState}, in the steady-state and low-intensity regime, the calculations of the ion-light interactions may be performed using the classical cross section. In particular, numerical simulations of the interaction in the GF can be performed in a similar manner as for light sources based on the inverse Compton scattering~\cite{Sprangle:1992zza}.  In such a case, one only needs to replace the Compton-scattering cross section with the absorption cross section given by Equation~\ref{eq:AbsorptionCrossSection}. We implement this approach using {\sf GF-CAIN} \cite{GF-CAIN}, a Monte Carlo event generator, being a customized (GF-adapted) version of the simulation code {\sf CAIN} \cite{CAIN} developed by K.\ Yokoya {\em et al.} at KEK-Tsukuba, Japan, for the ILC project \cite{ILC}. 

The PSI bunch is characterized in terms of standard high-energy particle beam parameters, such as emittances, beta-functions, etc., while the laser pulse is described by a space-time profile function, e.g.\ the Gaussian distribution.  Since the number of particles in a bunch can be huge, making the simulation of interactions of each individual particle with laser-photons unfeasible due to CPU-time and computer memory limitations, one usually replaces some number of physical particles by the, so-called, macroparticle and performs the actual simulations for a lower number of such macroparticles.  Then, to each macroparticle one assigns a Monte Carlo weight, which is equal to the number of physical particles  it represents.  Of course, the smaller the weight the better, because then the simulations are more precise in terms of systematic effects. 

The simulations proceed in such a way that at the beginning the bunch of macroparticles and a laser pulse are put some distance away in the $z$-direction in the LF, and then as time progresses they pass through each other in discrete time steps.  In each time step, the macroparticles interact with laser photons according to a given probability and, if excited, secondary photons are emitted from them through spontaneous emission.  After a predefined number of steps the simulation is finished.  Then, the outgoing particles can be transformed without interactions to a given value of the $z$-coordinate or the time-coordinate.  In this final position, the space-time coordinates and four-momenta of all outgoing particles are recorded. 

The basic quantity used in the simulation of the interaction of a single macroparticle $m$ ($m=1,\ldots, M$), being in the spatial position $\vec{r}$ and having the momentum $\vec{p}$ in the LF, with laser photons is the scattering probability during a time step $\Delta t_i = t_i - t_{i-1}$, where $t_i$ is the time of the $i$-th step, which is defined as 
\begin{equation}
    P_m(\vec{r},\vec{p},\vec{k},t_i) = \sigma_{\rm abs}(\vec{p},\vec{k})\left(1-\vec{\beta}\cdot\vec{k}/|\vec{k}|\right) n_p(\vec{r},\vec{k},t_i) c\Delta t_i\,.
    \label{eq:ScaProb}
\end{equation}
where $\vec{k}$ is the light wave vector, $\vec{p}$ and $\vec{\beta}$ are the PSI momentum and relativistic velocity, respectively, $n_p(\vec{r},\vec{k},t_i)$ is the local density of the photons, and $\sigma_{\rm abs}(\vec{p},\vec{k})$ is the absorption cross section given by \cite{Bessonov:1995eq}
\begin{equation}
    \sigma_{\rm abs}(\vec{p},\vec{k}) = \frac{\pi r_e c f\gamma_e}{[\gamma_L\omega^{\rm LF}(1-\beta\cos\psi) - \omega_0]^2 \,+ \,\gamma_e^2/4}\,,
    \label{eq:sigabs1}
\end{equation}
where $r_e$ is the classical electron radius, $f$ is the oscillator strength, $\omega^{\rm LF}$ is the irradiated light frequency in the LF, and $\psi$ is the angle between the directions of the light and PSI propagation in this frame.

The above can also be expressed in the following way
\begin{equation}
    \sigma_{\rm abs}(\vec{p},\vec{k}) = \frac{\sigma_0}{1 + 4\tau_e^2 \Delta\omega^2}\,,
    \label{eq:sigabs2}
\end{equation}
where 
\begin{equation}
    \sigma_0 = \frac{\lambda_0^2g_e}{2\pi g_g}\,,
\label{eq:sigma0}
\end{equation}
with $\tau_e=1/\gamma_e$ being the relaxation time of the excited state,  $g_g$ and $g_e$ being the degeneracy factors of the ground state $g$ and the excited state $e$, respectively, detuning $\Delta\omega = \omega - \omega_0$ measured in the IRF, where the IRF light frequency is given by $\omega=\gamma_L(1-\beta\cos\psi)\omega^{\rm LF}$, and $\lambda_0$ being the light central wavelength in the IRF.

For $P_m$ to act as probability, the size of the time step $\Delta t_i$ in Equation~\ref{eq:ScaProb} must be adjusted such that 
\begin{equation}
    0 \le P_m(\vec{r},\vec{p},\vec{k},t_i) \le 1, \; \forall_{m=1,\ldots M}\,.
\label{eq:ProbRange}
\end{equation}
In the simulations, the step size is set in such a way that if, for some macroparticle $m$ and time $t_i$, $P_m$ is larger than 1, the corresponding $\Delta t_i$ is divided into smaller steps until the condition given by Equation~\ref{eq:ProbRange} is fulfilled.

After computing the probability $P_m^i\equiv P_m(\vec{r},\vec{p},\vec{k},t_i)$, a scattering event is sampled using the (von Neumann) acceptance-rejection Monte Carlo method, see, for example, Ref.~\cite{Fishman:1996},
\begin{equation}
    \{0,1\} \ni \,  n_m^i = \int_0^1 dR\;\Theta(P_m^i - R)\,,
    \label{eq:RejMet}
\end{equation}
where $\Theta$ is the step function, i.e.\ a random number $R$ from the uniform distribution on $(0,1)$ is generated, and if $R\le P_m^i$, the event is accepted, otherwise it is rejected.  If the event is accepted, the corresponding macroparticle is marked as {\em excited}, which corresponds to the excitation of the PSI.  The macroparticle ``lives'' in the excited state for a time $\tau$ which is generated from the exponential distribution
\begin{equation}
    \zeta(\tau) = \frac{1}{\tau_e}\, e^{-\tau/\tau_e},\quad \tau \ge 0\,.
    \label{eq:ExpDis}
\end{equation}
While in the excited state, the macroparticle can interact with a laser photon and be deexcited by stimulated emission with the probability
\begin{equation}
    S_m(\vec{r},\vec{p},\vec{k},t_i) = \frac{g_g}{g_e}\,P_m(\vec{r},\vec{p},\vec{k},t_i)\,.
    \label{eq:StimEmProb}
\end{equation}
The stimulated emission event is generated, similarly as above, with the acceptance-rejection Monte Carlo method
\begin{equation}
    \{0,1\} \ni \, k_m^i = \int_0^1 dR\;\Theta(S_m^i - R)\,,
\label{eq:SERM}
\end{equation}
where $S_m^i\equiv S_m(\vec{r},\vec{p},\vec{k},t_i)$.  The corresponding emitted photon is not stored in the event record {as it goes along the laser pulse, and therefore does not reach a detector}.  If the excited ion is not deexcited by the stimulated emission within its lifetime $\tau$, it undergoes the spontaneous emission.  In such a case, the frequency $\omega_1$ as well as the polar $\theta_1$ and azimuthal $\phi_1$ angles of the emitted photon are generated in the ion reference frame (IRF), and then they are Lorentz-transformed to the LF.  The photon frequency is generated from the Lorentzian distribution, as given in Equation~\ref{eq:sigabs2}, while the emission angles are generated according to the angular distribution of fluorescence corresponding to a given atomic transition.  For such a photon, its LF four-momentum and space-time coordinates of the spontaneous emission are stored in the event record.

After the stimulated or spontaneous emission, the ion returns to its atomic ground state and is ready for absorption of another photon.  The whole above procedure is repeated for each macroparticle $m = 1,\ldots, M$ at a given time $t_i$ and is done for all time steps $\Delta t_i, i = 1,\ldots,I$.  The number of the spontaneously emitted photons from the PSI bunch is
\begin{equation}
    N_{\gamma} = \sum_{i=1}^I \sum_{m=1}^M \left(n_m^i - k_m^i\right) \frac{N_b}{M}\,,
    \label{eq:Ngam}
\end{equation}
where $N_b$ is the number of the PSI in the bunch, and $N_b/M$ is the Monte Carlo weight assigned to each macroparticle in the event record.

In a real experiment, the ions in the bunch and the photons in the laser pulse are not monoenergetic, but have some finite energy spread.  In the original {\sf CAIN} program the relative energy spread of particles in a bunch can be set in the input parameters and then the individual particle energy, i.e.\ the PSI Lorentz factor $\gamma_L$ in Equation~\ref{eq:sigabs1}, is generated from an appropriate Gaussian distribution.  On the other hand, the laser pulse  is assumed to be monochromatic.  For the inverse Compton scattering this is not important because the cross section does not depend strongly on the photon energy, but for the resonant absorption the finite energy spread of the laser pulse must be taken into account, particularly for the small linewidth $\gamma_e$ (cf.\ Equation~\ref{eq:sigabs1}).  It can be generated in the LF from the Gaussian distribution 
\begin{equation}
    \mathcal{D}(\omega^{\rm LF}) = \frac{1}{\sqrt{2\pi}\,\sigma_{\omega^{\rm LF}}}\,
    \exp\left[-\frac{(\omega^{\rm LF}-\omega_0^{\rm LF})^2}{2\sigma_{\omega^{\rm LF}}^2}\right]
    \label{eq:LasEspread}
\end{equation}
for a given relative frequency spread rms of a laser pulse $\sigma_{\omega^{\rm LF}}/\omega^{\rm LF}_0$, where $\omega^{\rm LF}_0$ is the central value of the laser-photon pulse frequency, adjusted to the central value of the absorption resonance for the central value of the PSI-bunch energy spread.
 
Alternatively, the photon frequency $\omega$ in IRF can be generated from the Lorentzian distribution of Equation~\ref{eq:sigabs2}, and then the scattering probability can be calculated by replacing the absorption cross section $\sigma_{\rm abs}(\vec{p},\vec{k})$ in Equation~\ref{eq:ScaProb} with the ``spread" cross section
\begin{equation}
    \sigma_{\rm spr}(\vec{p},\vec{k}) = \sigma_0\, \frac{\sqrt{\pi}\,\gamma_e}{2\sqrt{2}\,\sigma_{\omega}}\,
    \exp\left[-\frac{(\omega - \omega_0)^2}{2\sigma_{\omega}^2}\right]\,,
    \label{eq:GaussXsec}
\end{equation}
where $\sigma_{\omega}/\omega_0 = \sigma_{\omega^{\rm LF}}/\omega^{\rm LF}_0$ and $\sigma_0$ is given in Equation~\ref{eq:sigma0}.  In {\sf GF-GAIN} this method is used when $\gamma_e < 2\sqrt{2\ln 2}\,\sigma_{\omega}$,
which improves efficiency of event generation in such cases.  

{\sf GF-CAIN} has been cross-checked with the independent Monte Carlo generators {\sf GF-CMCC} and {\sf GF-Python}, and a good agreement with these programs has been found \cite{Curatolo:2018pza,GF-PoP-LoI:2019}. 

As discussed in the previous section, the above description can be applied reliably only for the H-like lead case of the GF presented in the third column of Table~\ref{table:OpticalPumpingParameters}.  Below we show some numerical results of the Monte Carlo simulations performed with {\sf GF-CAIN} for the input parameters given in the third column of Table~\ref{table:OpticalPumpingParameters}, with the supplementary parameters collected in Table~\ref{table:SimPars}.  The number of the spontaneously emitted photons from the PSI bunch is provided in the last row of Table~\ref{table:SimPars} -- it corresponds to the emission rate of $\sim 2.5$ of photons per ion.  It agrees within a factor of $1.65$ with the result of the semi-analytical calculations presented in the fourth column of Table~\ref{table:OpticalPumpingParameters}.  We have also checked that the emission rate grows linearly with the pulse energy between $0.05$ and $5\,$mJ.  
\begin{table*}[!htbp]
    \centering
    \caption{Some parameters for the {\sf GF-CAIN} simulations and the number of spontaneously emitted photons.}
	\begin{tabular}{lr}\hline\hline
		PSI beam  & ${}^{208}_{~82}\mathrm{Pb}^{81+}$\\
		\hline
		PSI mass $m$ & $193.687\,$GeV/c$^2$ \\
		PSI mean energy $\mathcal{E}$ & $578.9$ TeV \\
        Beta function at the interaction point $\beta_x=\beta_y$ & $0.5\,\mathrm{m}$\\
        Geometric emittance $\epsilon_x=\epsilon_y$ & $3\times 10^{-9}\;\mathrm{m\times rad}$\\
		\hline
		Laser (LF) & FEL (Gaussian)\\
		\hline
		Central wavelength of the laser in the LF $\lambda_0^{\rm LF}$ & $98.46\,$nm  \\
	    Rayleigh length $R_{L,x} = R_{L,y}$ & $7.5\,\mathrm{cm}$\\
		Interaction angle $\psi$ & 0$^\circ$\\
		\hline
 		 Atomic transition  & $1s \rightarrow 2p_{1/2}$\\
		\hline
		On-resonance absorption cross section $\sigma_0$ & $431.7\,$kb\\
		Angular distribution of emitted photons in the IRF $d^2p_1/(d\cos\theta_1 d\phi_1)$ & $1/(4\pi)$\\
		Maximum emitted photon energy in the LF $\hbar\omega_{\gamma}^{\rm max}$ & $450\,$MeV\\
		\hline
		Number of emitted photons per bunch $N_{\gamma}$ & $2.3\times 10^8$\\
		\hline\hline
	\end{tabular}	
\label{table:SimPars}	
\end{table*}

\begin{figure*}[!htbp]
    \centering
    \includegraphics[width=0.49\linewidth]{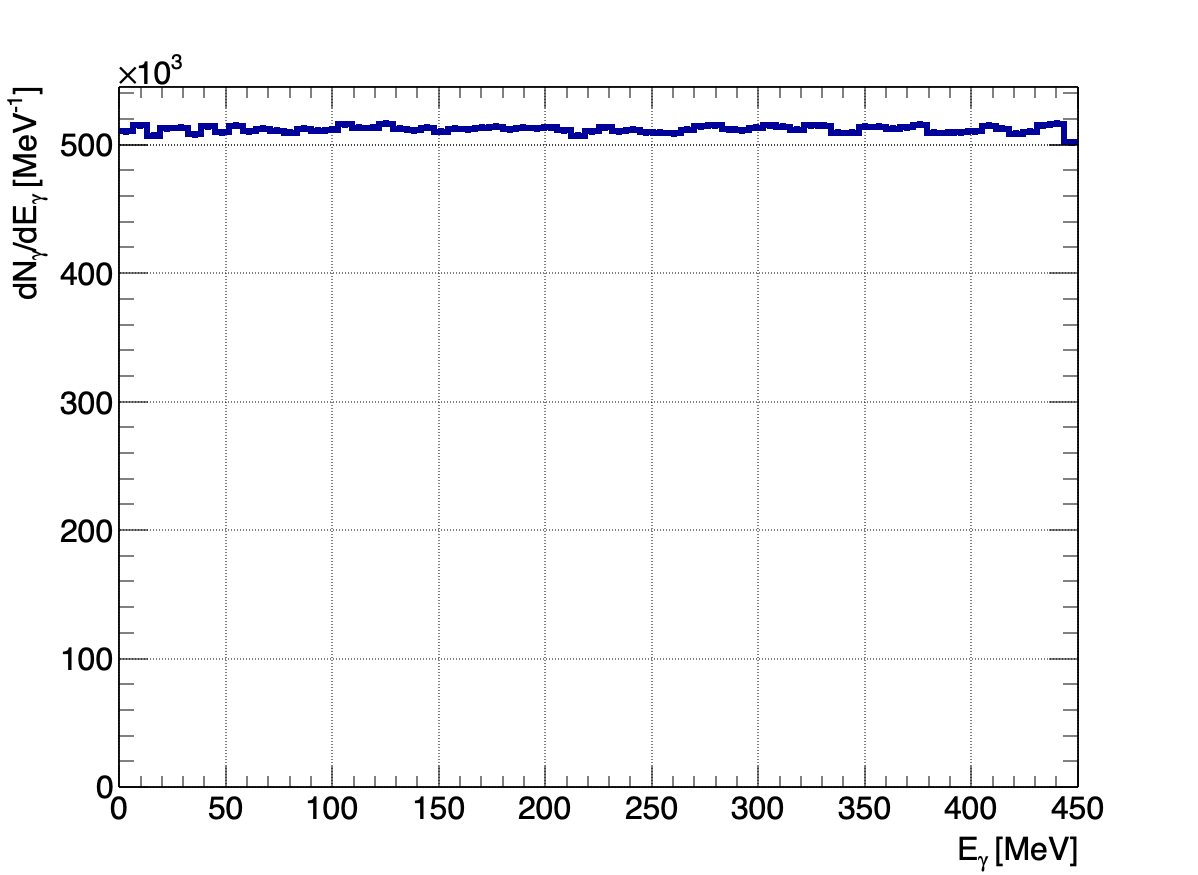}
    \includegraphics[width=0.49\linewidth]{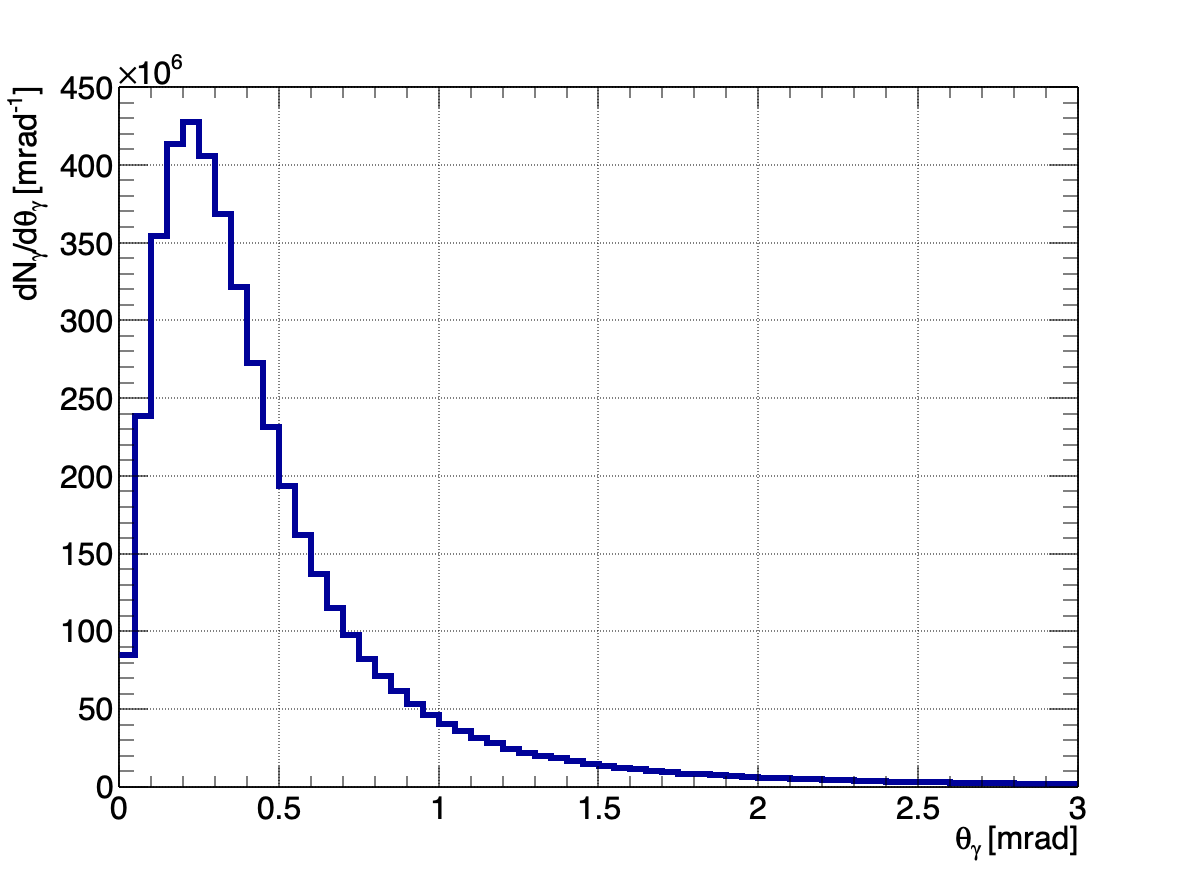}
    \caption{Distributions of energy (left) and polar angle (right) of the spontaneously emitted photons 
             from the H-like Pb bunch in the GF (in the LF).}
    \label{fig:EnergyAngle}
\end{figure*}

\begin{figure*}[!htbp]
    \centering
    \includegraphics[width=0.49\linewidth]{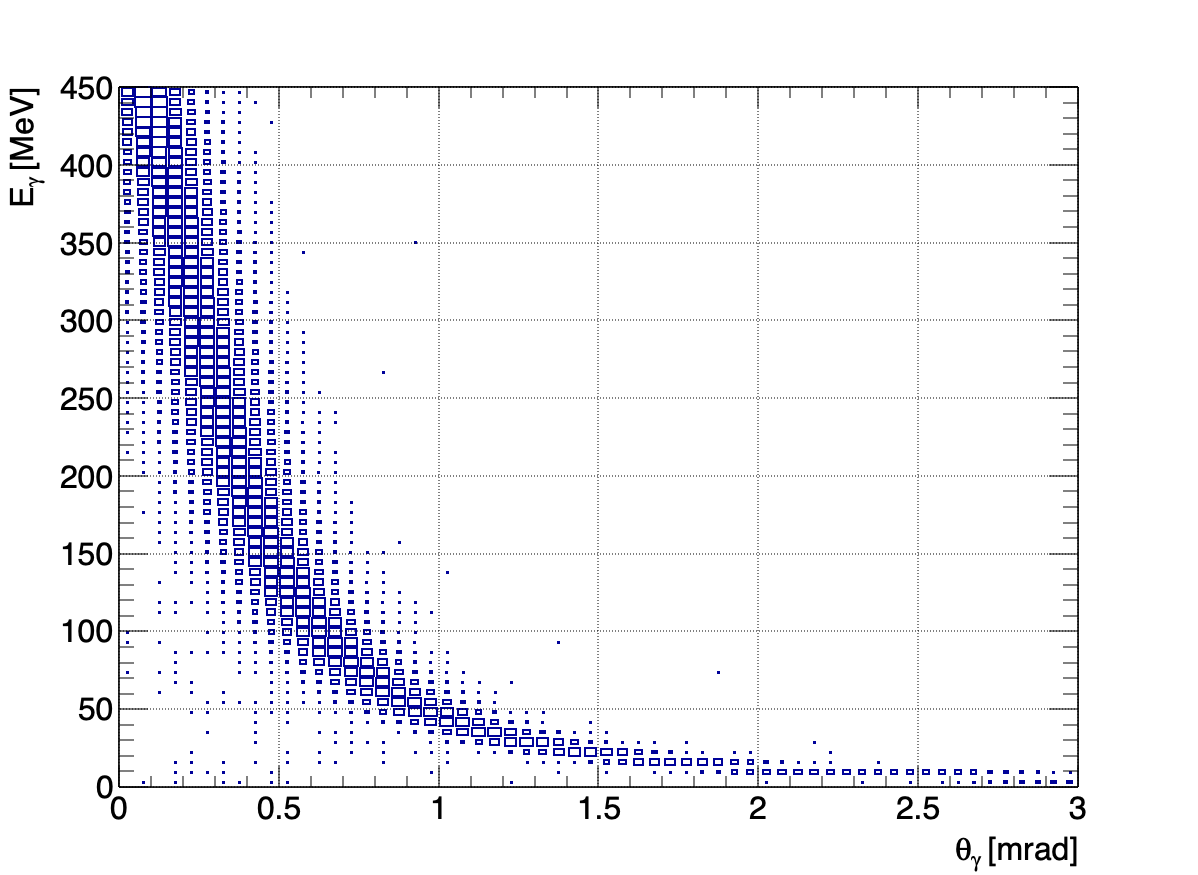}
    \includegraphics[width=0.49\linewidth]{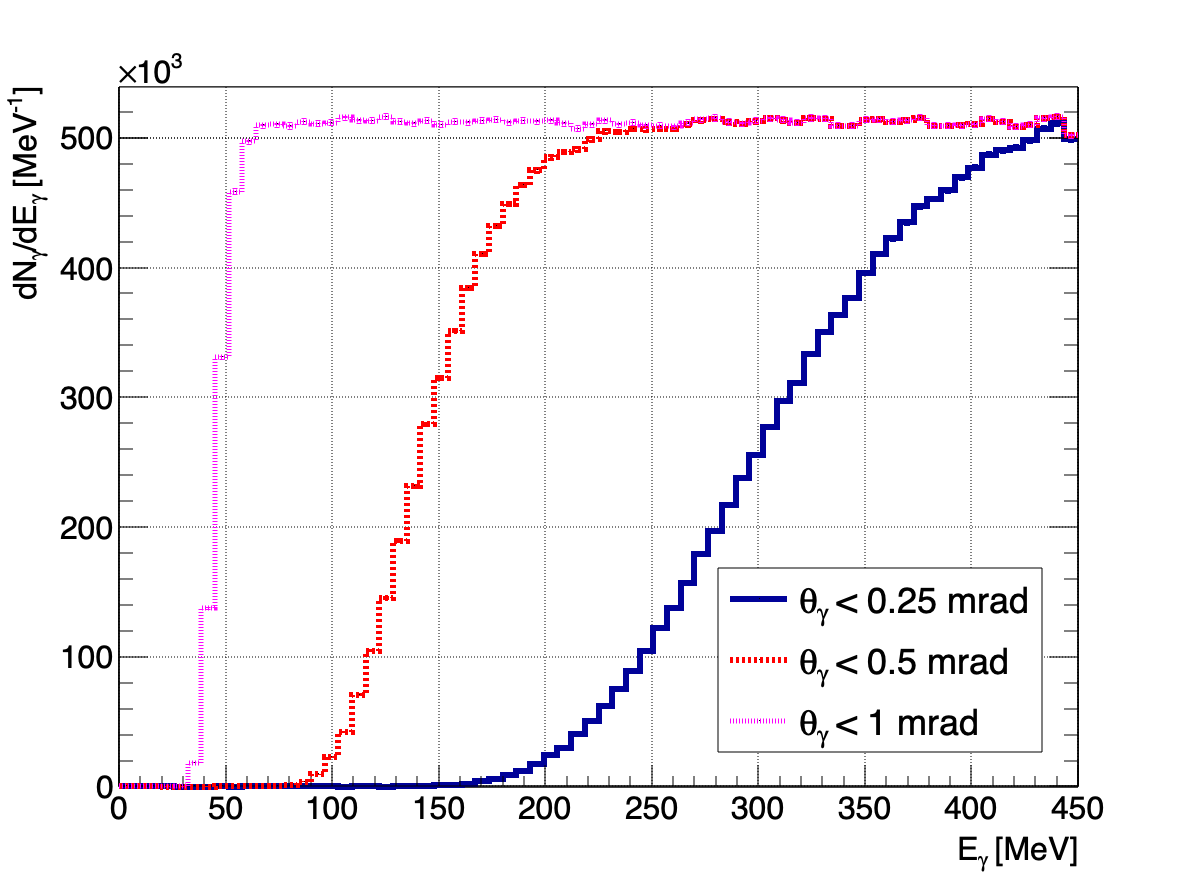}
    \caption{Distributions of energy versus polar angle (left) and energy for three angular upper cut-offs
             of the spontaneously emitted photons from the H-like Pb bunch in the GF (in the LF)}.
    \label{fig:EngThg}
\end{figure*}

The LF energy and emission-angle distributions of the outgoing photons are presented in Figures~\ref{fig:EnergyAngle} and~\ref{fig:EngThg}.  As discussed above, in the IRF the spontaneous emission is isotropic but its transformation to the LF modifies the emission pattern so that the emitted photons are strongly collimated.  As shown in Fig.~\ref{fig:EnergyAngle} (right panel), for a specific case of the system characterized with the parameters given in Table~\ref{table:SimPars}, most of the photons are emitted within an angle of $1\,$mrad, with the maximum at $\sim 0.25\,$mrad.  At the same time, the number of emitted photons versus the energy is characterized with the uniform distribution [Fig.~\ref{fig:EnergyAngle} (left panel)], which is a result of the isotropic emission in the IRF [$\phi_1\in {\cal U}(0,2\pi)$, $\cos\theta_1 \in {\cal U}(-1,1)$] and the Lorentz boost to the LF: $\omega^{\rm LF} = \gamma_L(1+\beta\cos\theta_1)\omega \Rightarrow dN_{\gamma}/d\omega^{\rm LF} \propto dN_{\gamma}/d\cos\theta_1$.  The difference in the distributions stems from nonlinear dependence between energy and angle, i.e.\ for higher energies, the same energy range corresponds to a smaller angular range than for lower energies, as can be seen in Figure~\ref{fig:EngThg} (left panel).

The left panel of Figure~\ref{fig:EngThg} shows strong correlations between the energy and angle of the emitted photons. Specifically, the most energetic photons are emitted at smallest angles.  In the right panel of Figure~\ref{fig:EngThg}, we present the photon energy distributions for three angular upper cut-offs: $0.25, 0.5$, and $1\,$mrad.  It shows that the emitted photon energy may be selected by applying angular collimators,
which can be in form of simple circular apertures placed some distance away from the interaction point, such that the sizes of the PSI-beam bunch and the laser pulse can be neglected.  

\section{Conclusions}
\label{sec:Concl}

We have evaluated and summarized the available literature data on spectroscopic properties of partially stripped ions (Li-like Pb, H-like Pb, He-like Ca and Li-like Ca) considered for the Gamma Factory project. We have demonstrated that lifetimes of relevant excited states of these ions are accurate to a percent level or better, while the corresponding transition energies reach between four and six digit accuracy. These numbers and their accuracies propagate into various parameters that will be necessary in experimental investigations.

Next, we have investigated ion--light interactions in the context of optical excitation, identifying two regimes determined by spectral properties of the ions in the bunch, i.e., the transition linewidth and the Doppler broadening, and the light pulse. We have shown that in the case of the Doppler broadening significantly exceeding the natural linewidth of the transition, efficient excitation needs to be based on Rabi oscillations and a good choice of the pulse total energy and length. We have theoretically motivated that under optimized (not experimental) conditions as much as $70\%$ of the atoms can be excited. We have also shown that in the case of the extremely short lifetimes of the excited state, the atom can undergo multiple excitation--deexcitation cycles, significantly increasing the number of photons emitted from the bunch. By investigating three ions of interest (Li-like ${}^{208}_{\phantom{0}82}$Pb$^{79+}$, Li-like ${}^{40}_{20}$Ca$^{17+}$, and H-like ${}^{208}_{\phantom{0}82}$Pb$^{81+}$), we calculated the efficiency of the excitation for a realistic set of parameters and hence estimated the number of photons scattered by the PSI bunches, but also identified the parameters optimizing the optical excitation and maximizing the number of scattered photons.

Finally, we have described Monte-Carlo simulations considering the process of the optical excitation of the H-like ${}^{208}_{\phantom{0}82}$Pb$^{81+}$ ion in a more realistic scenario.  For the simulations, we have developed the computer code {\sf GF-CAIN} which is currently limited to the steady-state regime, where the classical cross-section formulation can be applied.  Some exemplary results for the Gamma Factory realization at the LHC with the H-like Pb-ion beam have been presented.  The number of emitted photons agrees within a factor of about $1.5$ with semi-analytical calculations performed using the density-matrix formalism.  In our opinion, this agreement is satisfactory given the approximations employed in the latter calculations. 

The presented results are good starting point for experimental activities associated with the laser-light excitation of partially stripped ions in the Gamma Factory.  On the one hand, optical excitations of highly-charged heavy ions will allow to test theoretical calculations, providing access to such fundamental investigations as tests of quantum electrodynamics or violations of discrete symmetries~\cite{Budker2020Atomic}, but also offering means of spin polarization of PSI and studies of collisions of such ions.  On the other hand, the emission of secondary photons from the ions also offers schemes for generating extremely energetic (up to hundreds of MeV) and highly luminous ``light'' beams.  Due to their unique properties, such beams when extracted from production zones and collided with external targets can be used to produce high-intensity polarized electron, positron and muon beams, high-purity neutrino beams as well as high-flux neutron and radioactive-ion beams \cite{GF-PoP-LoI:2019,Placzek:2019xpw,GammaFactoryWorkingGroup:2020ely}.

\medskip
\textbf{Acknowledgements} \par 
SP acknowledges invaluable help and stimulating discussions with Dmitry Budker, Krzysztof Dzier\.z\c{e}ga, Alexey Petrenko, and Simon Rochester.  JB would like to thank Alexander Kramida and Andrey Surzhykov for extensive help in researching and computing spectroscopic data necessary for the Gamma Factory project.  WP acknowledges the fruitful collaboration with Camilla Curatolo.

\medskip

\bibliographystyle{MSP}
\bibliography{GammaFactory}

\end{document}